\documentclass[12pt]{article}
\usepackage{graphics}
\usepackage{epsfig}
\usepackage{latexsym}
\newcommand{\beq}{\begin{equation}}
\newcommand{\eeq}{\end{equation}}
\newcommand{\beqa}{\begin{eqnarray}}
\newcommand{\eeqa}{\end{eqnarray}}
\newcommand{\beqar}{\begin{eqnarray*}}
\newcommand{\eeqar}{\end{eqnarray*}}
\newcommand{\al}{\alpha}

\newcommand{\labell}[1]{\label{#1}} %{\label{#1}\qquad_{#1}} %
\newcommand{\reef}[1]{(\ref{#1})}
\newcommand\prt{\partial}

{\bf }

\newcommand\Tr{{\rm Tr}}

\def\no{{}^\circ \hspace{-0.16cm} {}_{\circ}}
\def\IR{{\hbox{{\rm I}\kern-.2em\hbox{\rm R}}}}

\begin{document}

\thispagestyle{empty}
\rightline{\small hep-th/0305177 \hfill MIT-CTP-3376}
\rightline{\small\hfill MCTP-03-26}
\vspace*{2cm}

\begin{center}
{ \LARGE The Rolling Tachyon as a Matrix Model}\\[.25em]
\vspace*{1cm}

Neil R. Constable$^1$ and Finn Larsen$^2$ 
\\
\vspace*{0.2cm}
${}^1${\it Center for Theoretical Physics, 
Massachusetts Institute of Technology }\\
{\it 77 Massachusetts Ave., Cambridge, MA-02139, USA}\\
\vspace*{0.1cm}
${}^2${\it Michigan Center for Theoretical Physics, the University of Michigan}\\
{\it 500 E. University, Ann Arbor, MI-48109, USA}
\vspace{0.2cm}

\vspace{2cm} ABSTRACT
\end{center}
We express all correlation functions in timelike boundary Liouville
theory as unitary matrix integrals and develop efficient techniques
to evaluate these integrals. We compute large classes of correlation 
functions explicitly, including an infinite number of terms in the boundary 
state of the rolling tachyon. The matrix integrals arising here also determine 
the correlation functions of gauge invariant operators in two dimensional 
Yang-Mills theory, suggesting an equivalence between the rolling tachyon 
and $QCD_2$ .

\vfill \setcounter{page}{0} \setcounter{footnote}{0}
\newpage

\section{Introduction}

The decay of unstable D--brane systems is a simple example of a time dependent
background in which one would like to understand the behaviour of string theory. 
More generally, the study of time dependent backgrounds is of interest for the
simple reason that we appear to live in one. Unfortunately this is a notoriously difficult issue
to even formulate in a clean way. It is therefore of no small interest that the rolling tachyon 
backgrounds of Sen~\cite{rolling} are described in terms of exactly soluble boundary 
conformal field theories. These backgrounds are therefore tractable, and one may 
hope that lessons learned here carry over to more general time dependent
situations. 

The time dependent background studied in this paper is the rolling tachyon corresponding 
to the decay of a D25 brane in bosonic string theory. We will be specifically interested in 
the case where the tachyon on the worldvolume of the D25-brane sits at the top of its potential 
at $t=-\infty$ and rolls to the minimum as $t\rightarrow +\infty$~\cite{opensbrane,lnt,GS,sthermo}. 
In terms of conformal field theory this is described by the usual $c=25$ worldsheet theory for the 
spatial fluctuations of the open string plus the action
\beq
-\frac{1}{2\pi}\int_{\Sigma}\prt X^{0}\bar{\prt}X^{0} +g\int_{\prt\Sigma}e^{X^0(e^{ix^0})}~,
\labell{TBL}
\eeq
for the temporal fields. We will refer to this theory of a negative norm boson with an 
exponential boundary interaction as Timelike Boundary Liouville theory and resist the terminology 
${1\over 2}$S-brane. Many other interesting avenues of investigation into rolling tachyons have 
been pursued in the recent literature~\cite{moresen,sbranes,moresbrane,minahan,terashima,okuda,rey1,DaC,
Kut}.

This paper will focus on general classes of correlation functions in 
Timelike Boundary Liouville theory (TBL).
The main result of the investigation is a demonstration that all correlators in this
theory, and hence the rolling tachyon background, can be expressed as unitary matrix 
integrals. Moreover, these integrals permit an expansion in simple quantities
from the theory of groups, such as the characters of the symmetric groups.
These relations provide an efficient, purely algebraic, algorithm for computing
all correlation functions of the theory. As explicit examples we 
obtain infinitely many coefficients in the boundary state of the rolling tachyon. 
We also compute $m$-point correlators of exponentials of the field $X^0$ in the background 
of the rolling tachyon, for bulk and boundary fields.

The appearance of matrix integrals strongly suggests that 
the full dynamics of this time dependent background should 
be captured by a unitary matrix model. 
The obvious question is, which matrix model?
An indication of which class of matrix models we should look at comes from the fact
that correlation functions of timelike boundary Liouville 
theory, when expressed in terms of $U(n)$ matrices, are 
easily recognizable as correlation functions of gauge invariant 
operators in two-dimensional Yang-Mills
theory~\cite{wati1,wati2}. In particular, the one point functions which lead to the boundary state 
coefficients are precisely the same as the correlation functions between pairs of Wilson
loops in $QCD_2$, in the limit of small separations. We are thus led to conjecture that the timelike 
boundary Liouville theory is the same theory as $QCD_2$. 
As it is widely believed that two dimensional Yang-Mills can itself be formulated as a matrix model, 
the same will be true for TBL. The matrix model description would amount to a holographic 
projection of TBL to one dimension less. The surprising twist to the story is that the holographic
description itself is the gauge fixed version of $QCD_2$, {\it i.e.} a theory in one dimension {\it more}. 
It would clearly be interesting to develop these relations further. Recently other relations
between matrix models and tachyons have been proposed in~\cite{mv}. 

%Schematically the equivalence can be summarized as,
%\beqa
%\langle\prt X^0(0)\bar{\prt}X^0(0)\rangle_{TBL} &\leftrightarrow & \langle\Tr(U^{\dagger })\Tr(U)\rangle_{Matrix} \\ \nonumber
%\langle e^{X^0(z,\bar{z})}\rangle_{TBL} &\leftrightarrow & \langle\det(1-zU^{\dagger})\det(1-\bar{z}U)\rangle_{Matrix} 
%\\ \nonumber
%\int\prod_i\frac{dt_i}{2\pi}\langle \prod_ie^{X^0(e^{it_i})}\rangle &\leftrightarrow & \int dU
%\labell{identifications}
%\eeqa
%where $z,\bar{z}$ are coordinates on the interior of the disk and the 
%variables $t_i$ represent points on the boundary of the disk. 

The remainder of the paper is organized as follows. In section two we review
the basic observation of~\cite{lnt}
that led to our investigation, namely, that the partition function of the rolling tachyon 
background is related to the Haar measure of the unitary group. We also review how 
$S_N$, the symmetric group of $N$ objects, encodes the spectrum of the closed bosonic 
string. In section three we show that bulk one point functions of massive closed string states are in fact matrix 
integrals. We establish our technique for evaluating these integrals and obtain general 
expressions for infinitely many boundary state coefficients. In section four we consider
bulk $m$-point functions, demonstrating how these may also be written as matrix integrals 
and readily evaluated.  These results are extended to
boundary correlators in section five. In section six we discuss our
conjecture relating timelike boundary Liouville theory to 
two dimensional gauge theory and hence a matrix model. 
Finally, in an attempt at making the paper somewhat self-contained, we have
included several appendices reviewing some facts from the representation theory of
the unitary groups, the characters of the symmetric group, 
and the theory of symmetric functions.

\section{Preliminaries}
In this section we review a few of the key ingredients needed in this paper.
First, the representation of correlators in the rolling tachyon background as integrals over 
$U(n)$ group manifolds; and then the connection between string states and conjugacy 
classes of the symmetric group. 

\subsection{Correlators as $U(n)$ Integrals}
The basic observables in the rolling tachyon background are the correlation functions of vertex 
operators, corresponding to open and closed string states. 
A general correlator is of the form
\beq
{\cal{A}}^{(\Pi V_i)} = \langle\prod_iV_i(z,\bar{z})\rangle_{TBL} = 
\langle\prod_iV_i(z,\bar{z})e^{-I_{bndy}}\rangle~,
\labell{basic}
\eeq
where $z,\bar{z}$ are coordinates on the unit disk and the boundary interaction is 
that of the rolling tachyon
\beq
I_{bndy} = g\int dt ~\exp(X^0(t))~. 
\labell{bact}
\eeq
One approach to evaluating these expressions is to treat the boundary interaction perturbatively and write
\beq
{\cal{A}}^{(\Pi V_i)}  = \sum_{n=0}^{\infty}\frac{(-2\pi g e^{x^0})^n}{n!}
\langle\prod_{i}V_i(z,\bar{z})\int\prod_{i=1}^{n}\frac{dt_i}{2\pi}e^{{\hat X}^0(e^{it_i})}\rangle
\labell{perturb}~,
\eeq
where the field $X^0$ was divided into a zeromode $x^0$ and a fluctuating part ${\hat X}^0$. 
We will leave the zero-mode $x^0$ unintegrated, a standard procedure when interpreting 
the CFT in spacetime as a rolling tachyon. It is useful to introduce a separate notation for 
the $n$th order contribution to the correlator eq. (\ref{perturb}) 
\beq
{\cal{A}}^{(\Pi V_i)}_n \equiv {1\over n!}
\langle\prod_{i}V_i(z,\bar{z})\int\prod_{i=1}^{n}\frac{dt_i}{2\pi}e^{{\hat X}^0(e^{it_i})}\rangle~,
\eeq
and so write
\beq
{\cal{A}}^{(\Pi V_i)} = \sum_{n=0}^\infty (-\tilde{g})^n {\cal{A}}^{(\Pi V_i)}_n~,
\eeq
where $\tilde{g}\equiv 2\pi g e^{x^0}$.

The key observation for the techniques developed in this paper is that the contractions
that do not involve the vertex operators take the form
\beq
\langle\prod_{i=1}^{n}e^{{\hat X}^0(e^{it_i})}\rangle = \prod_{i<j} |e^{it_i}-e^{it_j}|^2 =
\prod_{i<j} 4\sin^2\left(\frac{t_i-t_j}{2}\right)\equiv \Delta^2(t)~,
\labell{vander}
\eeq
where  $\Delta(t)$ is the Vandermonde determinant for the group $U(n)$~\cite{lnt}. 
It follows that the disk amplitude with no vertex operator insertions  becomes
\beq
{\cal A}^{\rm vac}_n={1\over n!}
\int\langle\prod_{i=1}^{\infty}\frac{dt_i}{2\pi}e^{{\hat X}^0(e^{it_i})}\rangle = 
{1\over n!}\int\prod_{i=1}^{n}\frac{dt_i}{2\pi}\Delta^2(t) = \frac{1}{{\rm vol}(U(n))}\int dU = 1~,
\labell{haar}
\eeq
where $dU$ is the Haar measure for $U(n)$. Summing over all orders in the perturbation series,
we find 
\beq
{\cal A}^{\rm vac}=\sum_{n=0}^\infty (-\tilde{g})^n=\frac{1}{1+{\tilde g}} \equiv f(x^0)~.
\labell{fdef}
\eeq
The function $f$ is the partition function of the theory, treated as a function
of the unintegrated zero-mode.

The appearance of the Vandermonde determinant in the vacuum amplitude begs the
question of whether other correlation functions, with the $V_i(z,{\bar z})$ retained, can be similarly 
represented in terms of unitary matrices. We will see that this is indeed the case. 
  
\subsection{Closed Strings and the Symmetric Group}
The mass spectrum of closed bosonic string theory is given by
\beq
{1\over 2}m^2 +2 = \sum_{n}N_nn + 
\sum_{\tilde{n}}\tilde{N}_{\tilde{n}}\tilde{n} \equiv N +\tilde{N}~,
\label{spec}
\eeq
in units with $\al^{\prime} = 1$. The level matching condition states 
that $N=\tilde{N}$, while the $N_n$ and $\tilde{N}_{\tilde{n}}$ need not be related. The spectrum
at a given mass level is thus labelled by a pair of partitions of the integer $N$. 
For example, if ${1\over 2}m^2+2 = 3 + 3$ the possible string states are identified by 
pairs of partitions of $3$. The partitions are $(3),(1,2),(1,1,1)$ in this case. The state 
\beq
\al_{-3}\tilde{\al}_{-2}\tilde{\al}_{-1}|0\rangle~,
\label{examp}
\eeq
corresponds to the partition $(3)$ for the left movers and the partition $(1,2)$ for the right movers.
The states for other pairs of partitions are readily written down as well.

The utility in taking this point of view is that the partitions of $N$ are in a one to one correspondence
with the conjugacy classes of the symmetric group $S_N$. So we may just as well view the oscillator 
structure of each `side' of the string as being labelled by a conjugacy class of $S_N$. 
In the example above we are dealing with $S_3$ and the state written in eq.~\reef{examp}
corresponds to choosing the conjugacy class of long cycles  $(123)$ for the 
left movers and the conjugacy class  of $(12)(3)$ for the right movers. 
Here we are employing the standard notation for elements of $S_3$. 
In general, the partition $N=N_1\nu_1 + N_2\nu_2 +\cdots N_k\nu_k$, 
where the $N_i$ are the multiplicities of the $\nu_i$, corresponds to the oscillators
\beq
\al_{-\nu_1}^{N_1}\al_{-\nu_2}^{N_2}\cdots\al_{-\nu_k}^{N_k}~. 
\eeq
This set of oscillators is thus labelled by the conjugacy class of $S_N$ which contains $N_1$ 
cycles of length $\nu_1$, $N_2$ cycles of length $\nu_2$ and so on. We will denote
this conjugacy class $\sigma=(\nu_1^{N_1},\nu_2^{N_2},\cdots,\nu_k^{N_k})$. Here and in the folowing all oscillators
$\alpha_{\nu_i}$ will be assumed to be temporal unless indicated otherwise.

A general massive string state at level $N= {1\over 4}m^2+1$  can be expressed in
oscillator notation as
\beq
{\cal N}^{(\sigma;\tilde{\sigma})}\al_{-\nu_1}^{N_1}\al_{-\nu_2}^{N_2}\cdots\al_{-{\nu_k}}^{N_k}
\tilde{\al}_{-\tilde{\nu}_1}^{\tilde{N}_1}\tilde{\al}_{-\tilde{\nu}_2}^{\tilde{N}_2}
\cdots\tilde{\al}_{-\tilde{\nu}_{\tilde{k}}}^{\tilde{N}_{\tilde{k}}}|0,0\rangle
\equiv |\sigma,\tilde{\sigma}\rangle~,
\label{massstate}
\eeq
where $\sigma,\tilde{\sigma}$ denote the conjugacy classes of $S_N$ which label the state we are
considering and
\beq
{\cal N}^{(\sigma;\tilde{\sigma})} = \left[ \prod_i\nu_i^{N_i}N_i!\prod_{\tilde{i}}\tilde{\nu}_{\tilde{i}}^
{\tilde{N}_{\tilde{i}}}
\tilde{N}_{\tilde{i}}! \right]^{-{1\over 2}}~,
\labell{norm}
\eeq
is the normalization factor. The corresponding vertex operators are given by
\beq
V^{(\sigma;\tilde{\sigma})}(z,{\bar z}) = {\cal N}^{(\sigma;\tilde{\sigma})}
\prod_{i=1}^{k}
\left(\frac{\sqrt{2}}{(\nu_i-1)!}\partial^{\nu_i}X^0(z,{\bar z})\right)^{N_i}
\prod_{\tilde{i}=1}^{\tilde{k}}
\left(\frac{\sqrt{2}}{(\tilde{\nu}_{\tilde{i}}-1)!}\bar{\partial}^{\tilde{\nu}_{\tilde{i}}}X^0(z,{\bar z})
 \right) ^{\tilde{N_{\tilde{i}}}}~,
\label{vertexop}
\eeq
where the numerical constants were determined using the operator-state 
correspondence (as in~\cite{polchinski}). 

\section{Couplings to Closed Strings}
In this section we introduce the matrix method by considering the simplest 
correlators, the one point functions of closed strings in the rolling tachyon
background. We also point out a suggestive connection to $QCD_2$.
Finally, we relate our results to those obtained using boundary state methods. 

\subsection{Correlators as Matrix Integrals}

The $n$th order contribution to the disk amplitude with a single closed string inserted at the origin,
can be written as
\beq
{\cal A}_{n}^{(\sigma,\tilde{\sigma})}=\int\prod_{i=1}^n\frac{dt_i}{2\pi}A_n^{(\sigma;\tilde{\sigma})}=
\int\prod_{i=1}^n\frac{dt_i}{2\pi} \langle\no V^{(\sigma;\tilde{\sigma})}(0,0)\no 
\prod_{l=1}^{n}\exp(X^0(w_l))\rangle~,
\label{amp}
\eeq
where the $w_l=e^{it_l}$ represent the positions of the tachyon vertex operators associated with the
rolling tachyon background, and the notation $\no~~\no$ denotes boundary normal ordering. 
%and  will be
%taken to lie on on the boundary of the disc at the end of the calculation. 
The Green's function on the unit disk is
\beq
G_l \equiv G(z,w_l) = \log|z-w_l|+  \log|z\bar{w}_l-1|~,
\label{green2}
\eeq
for a field with temporal signature. Carrying out contractions involving the
closed string vertex we find
\beq
A^{(\sigma;\tilde{\sigma})}_{n} 
= {\cal N}^{(\sigma;\tilde{\sigma})}
\prod_{i=1}^{k}
\left[ {\sqrt{2}\over (\nu_i-1)!}\sum_{l=1}^{n}\partial^{\nu_i}G_l
\right]^{N_i}\prod_{\tilde{i}=1}^{\tilde{k}}\left[{\sqrt{2}\over (\tilde{\nu}_{\tilde{i}}-1)!}\sum_{\tilde{l}=1}^{n}
\bar{\partial}^{\tilde{\nu}_{\tilde{i}}}
G_{\tilde{l}}\right]^{\tilde{N}_{\tilde{i}}}\langle\prod_{j=1}^{n}\exp(X^0(w_j))\rangle~.
\label{amp2}
\eeq
It is straightforward to show that
\beq
\partial^{\nu_i}G_l = - (\nu_i-1)! \exp(-i\nu_it_l)~,
\eeq
where we have taken $z=0$ and $w_l = e^{it_l}$. The factor $(\nu_i-1)!$ cancels a
similar factor in the vertex operator normalization; and so, after using eq.(\ref{vander}), we find
\beq
A^{(\sigma;\tilde{\sigma})}_{n} ={\cal N}^{\prime(\sigma;\tilde{\sigma})} ~\Delta(t)^2~\prod_{i=1}^{k}
\left[\sum_{l=1}^{n}e^{-it_l\nu_i}\right]^{N_i}
\prod_{\tilde{i}=1}^{\tilde{k}}\left[\sum_{\tilde{l}=1}^{n}
e^{it_{\tilde{l}}\tilde{\nu}_{\tilde{i}}}\right]^{\tilde{N}_{\tilde{i}}}~,
\label{ampfinal}
\eeq
where
\beq
{\cal N}^{\prime(\sigma;\tilde{\sigma})}=
\left[ \prod_i({\nu_i\over 2})^{N_i}N_i!\prod_{\tilde{i}}({\tilde{\nu}_{\tilde{i}}\over 2})^{\tilde{N}_{\tilde{i}}}
\tilde{N}_{\tilde{i}}! \right]^{-{1\over 2}}~.
\eeq
This generalizes the results of~\cite{lnt} to the case of the most 
general massive closed string.

%Using eqns.~\reef{vander} and \reef{ampfinal} we notice that \reef{amp} can be rewritten as,
%\beq
%I^{\sigma\tilde{\sigma}}_{n} = \int\prod_{i=1}^{n}\frac{dt_i}{2\pi}\Delta^2(t)
%\prod_{i=1}^{k}\left[\sum_{l=1}^{n}e^{-it_l\nu_i}\right]^{N_i}
%\prod_{\tilde{i}=1}^{\tilde{k}}\left[\sum_{\tilde{l}=1}^{n}
%e^{it_{\tilde{l}}\tilde{\nu}_{\tilde{i}}}\right]^{\tilde{N}_{\tilde{i}}}
%\label{eigenint}
%\eeq
Our goal is to represent the amplitude eq.(\ref{amp}) as a matrix integral.
A $U(n)$ matrix $U$ can always be written as $U=\exp(iT)$ for some Hermitean 
matrix $T$. Further the eigenvalues of $U$ can be written as $e^{it_i}$ where 
the $t_i$ are the eigenvalues of $T$. In this diagonal basis we see that
\beq
\Tr(U) = \sum_{i=1}^{n}\exp(it_i)~,
\eeq
or more generally,
\beq 
\left[\Tr(U^{\nu_i})\right]^{N_i} = \left[\sum_{l=1}^{n}\exp(it_l\nu_i)\right]^{N_i}~.
\label{trace}
\eeq 
The amplitude eq.(\ref{amp}) with integrand eq.(\ref{ampfinal}) can therefore
be written as
\beq
{\cal A}_n^{(\sigma;\tilde{\sigma})} = {\cal N}^{\prime(\sigma;\tilde{\sigma})} I_n^{(\sigma;\tilde{\sigma})}~, 
\eeq
where
\beq
 I_n^{(\sigma;\tilde{\sigma})} =
\frac{1}{{\rm vol}(U(n))}\int dU \prod_{i=1}^{k}\left[Tr(U^{\nu_i})\right]^{N_i}
\prod_{\tilde{i}=1}^{\tilde{k}}
\left[Tr(U^{\dagger \tilde{\nu_{\tilde{i}}}})\right]^{\tilde{N}_{\tilde{i}}}~.
\label{matint}
\eeq
This is a general expression for {\it all} one point functions of massive string modes  
in the rolling tachyon background. Integrals of this form  can be evaluated elegantly and 
efficiently, by exploiting group theory methods. 

Before showing how to do this it is worth noting the intriguing 
connection to two-dimensional $U(n)$ gauge theory. Namely, the matrix integral in 
eq.~\reef{matint} is {\it exactly} the expression one 
obtains in $QCD_2$ as the correlation function between a pair of Wilson loops in the limit 
where the area of the Riemann surface (target space)
shrinks to zero~\cite{wati1,wati2}. The precise connection between the rolling tachyon 
background and $QCD_2$ is a bit 
mysterious, since in the case of interest here each term in the perturbative expansion~\reef{perturb} 
corresponds to a Wilson loop
correlator in a different gauge theory {\it i.e.} $U(2),U(3)$ and so on. We will comment more
on this fascinating connection in section six.

\subsection{Integrals of Class Functions}
The tool needed to evaluate the amplitudes given in eq.\reef{matint} is a classic 
piece of mathematics known as Schur-Weyl duality, which relates the irreducible representations of $U(n)$ 
to those of the symmetric group (See Appendix A for more details). 

Consider a function of the form $f(U) =\prod_{i=1}^{k}\left[Tr(U^{\nu_i})\right]^{N_i} $, with 
$U$ in the defining $n\times n$ representation of $U(n)$. $f(U)$ is invariant under 
conjugation of $U$ by elements of $U(n)$; and so depends only on the conjugacy classes
of $U$. Functions with this property are known as class functions. The integrand in eq.~\reef{matint} is the 
product of class functions on $U(n)$ and this makes the evaluation 
of the integral extremely easy. This is because the characters $\chi_{\lambda}(U)$ of the irreducible 
representations of $U(n)$ provide an orthonormal basis for class functions.
%\footnote{Recall that the 
%characters of an abstract group element $U$ are given by 
%$\chi_{\lambda}(U)=\Tr_{\lambda}(U)$ where the notation indicates that the trace is taken in 
%the irreducible representation labelled by $\lambda$.}. 
In the present case  $f(U)$ can be expanded in this basis 
as
\beq
f(U)= \prod_{i=1}^{k}[Tr(U^{\nu_i})]^{N_i} =  \sum_{\lambda\leq n}\chi_\lambda (\sigma)\chi_{\lambda}(U)~,
\labell{basisexp}
\eeq
where the summation index $\lambda$ refers to irreducible representations of the 
symmetric group $S_N$; and $\chi_{\lambda}(\sigma)$ is the character of $\sigma\in S_N$ in the
irreducible representation $\lambda$. Recall that $\sigma$ is a representative of the conjugacy class 
which labels the left-moving side of the string state. The irreducible representations of $S_N$ 
are classified by the partitions of $N$ which, in turn, we can picture as Young frames with $N$ boxes. 
Of course, a given Young frame also corresponds to an irreducible representation of $U(n)$, and 
it is in this sense that the index $\lambda$ is used in eq.\reef{basisexp}  to label $U(n)$ characters, 
as well as representations of $S_N$. However, in some cases Young frames with $N$ boxes have 
more than $n$ boxes in one column, and then the corresponding $U(n)$ representation vanishes,
by complete anti-symmetry of the corresponding tensors. This imposes an important restriction on 
the sum over $\lambda$ in eq.\reef{basisexp}. We have introduced the notation $\lambda\leq n$ 
as a shorthand to remember that we should sum only over $S_N$ representations that
make sense also as $U(n)$ representations.

We are now ready to use the expansion eq.\reef{basisexp} to evaluate the matrix integral
eq.~\reef{matint}. We find
\beqa
I^{(\sigma;\tilde{\sigma})}_{n}
&=& \frac{1}{{\rm vol}(U(n))}
\int dU  \prod_{i=1}^{k}\left[Tr(U^{\nu_i})\right]^{N_i}
\prod_{\tilde{i}=1}^{\tilde{k}}
\left[Tr(U^{\dagger \tilde{\nu_{\tilde{i}}}})\right]^{\tilde{N}_{\tilde{i}}}\nonumber
\\ \nonumber
&=&\frac{1}{{\rm vol}(U(n))}
\sum_{\lambda\lambda^{\prime}\leq n}\chi_\lambda(\sigma)\chi_{\lambda^{\prime}}(\tilde{\sigma})
\int dU\chi_{\lambda}(U)\chi_{\lambda^{\prime}}(U^{\dagger}) \\ \nonumber
&=& \sum_{\lambda\lambda^{\prime}\leq n}\chi_\lambda(\sigma)\chi_{\lambda^{\prime}}(\tilde{\sigma})
\delta_{\lambda\lambda^{\prime}} \\ 
&=& \sum_{\lambda\leq n}\chi_{\lambda}(\sigma)\chi_{\lambda}(\tilde{\sigma})~,
\label{integral}
\eeqa
where, in going from the second to third line, we have used 
the orthogonality of group characters
\beq
\frac{1}{{\rm vol}(U(n))}\int dU \chi_{\lambda}(U)\bar{\chi}_{\lambda^{\prime}}(U) = 
\delta_{\lambda\lambda^{\prime}}~,
\labell{orthochar}
\eeq
and also $\chi_{\lambda}(U^\dagger) = \bar{\chi}_{\lambda}(U)$ for unitary representations. The general 
amplitude thus reduces to evaluating a sum over the characters of the irreducible 
representations of $S_N$. Before considering explicit examples it is worthwhile to make a few 
general remarks.

When $n\ge N$ there are no Young frames with $N$ boxes that have more than $n$ columns;
so, in this case, the final sum in eq.\reef{integral} is over all irreducible representations of $S_N$ and is 
given by the completeness relation\footnote{see for example ref.~\cite{reptheory}}
\beq
I^{(\sigma;\tilde{\sigma})}_{n} =\sum_{\lambda}\chi_{\lambda}(\sigma)\chi_{\lambda}(\tilde{\sigma}) = 
\delta_{\sigma\tilde{\sigma}}\prod_k\nu_k^{N_k}N_k!~~~~~;~~n\ge N~,
\labell{sum1}
\eeq
where the integers $\nu_k,N_k$ are those defining the conjugacy class $\sigma$. 

When $n<N$ the sum in eqn.~\reef{basisexp} does not run over all irreducible representations of $S_N$ and 
there is no simple expression analogous to eqn.~\reef{sum1}. In particular, characters 
of different conjugacy classes are not in general orthogonal with the restricted sum. In these cases the 
final summation must be performed by directly evaluating the characters of the representations 
which do appear. For all the states at level $N$ what we need is simply the character table of $S_N$,
a standard quantity which can be computed using a variety of techniques. In 
Appendix B we outline one such technique, known as the Murnaghan-Nakayama Rule. 

To compute the amplitude for a given closed string state we want to sum over {\it all} orders in 
perturbation theory, {\it i.e.} over all values of $n$. The cases $n\ge N$ and $n<N$ therefore
both arise, for any amplitude. As we will show next, the matrix method is nevertheless a practical, 
indeed efficient, method to compute couplings to closed strings.

\subsection{Examples}

%For example let us return to the discussion in section 3.1. In that
%case we had $M=3$. The summation over all Young frames in 
%eqn.~\reef{basisexp} is then simply the sum over
%the diagrams in figure~\reef{figY1}.
%\begin{figure}[ht!]
%\center{\includegraphics{young1.eps}}
%\caption{Young Frames for $S_3$}
%\label{figY1}
%\end{figure}
%Each Young frame has the dual interpretation as defining both an irreducible 
%representation of $S_3$ 
%and an irreducible representation of $U(n)$.
%Note that when $n=2$ the first diagram in figure~\reef{figY1} does not 
%contribute to the sum since there are 
%no representations of $U(n)$ with more than $n$-boxes in a given column. 
%Likewise when $n=1$ neither the 
%first or second diagram appears in the summation. For $n\ge M$ all irreducible 
%representations of $S_M$ appear
%in the decompostion. 

We are now in a position to evaluate the amplitudes
\beq
{\cal A}^{(\sigma;\tilde{\sigma})} = \langle \no V^{(\sigma;\tilde{\sigma})}\no\rangle_{TBL} ~,
\labell{finalans1}
\eeq
for arbitrary closed string states. We first compute the quantities,
\beq
I^{(\sigma;\tilde{\sigma})} =\sum_{n=0}^\infty (-{\tilde g})^n I^{(\sigma;\tilde{\sigma})}_n~,
\eeq
and then multiply by the overall normalization factor to find ${\cal A}^{(\sigma;\tilde{\sigma})} 
={\cal N}^{\prime(\sigma;\tilde{\sigma})} I^{(\sigma;\tilde{\sigma})}$.

As the simplest example of a state at level $N$ consider 
\beq
{1\over N}\al_{-N}\tilde{\al}_{-N}|0\rangle~.
\labell{ex1}
\eeq
This corresponds to choosing the conjugacy class of long cycles $(12\cdots N)$ in 
$S_N$ to label both the right and left movers. In terms of partitions of $N$, the states correspond to the 
trivial partition, namely $\nu_1=N,\, N_1 = 1$, {\it i.e.} $\sigma={\tilde\sigma}=(N)$. 
For $n\ge N$, the orthogonality relation eq.~\reef{sum1}  immediately gives 
\beq
I^{(N;N)}_n = N  \,\,\,\,\,\,\,\ ;~~ n\ge N~.
\labell{longsimp}
\eeq
The case of $n<N$ is only slightly more difficult. As discussed above one must sum the 
characters of irreducible representations of $S_N$ which correspond to Young frames 
which do not contain any columns with more than $n$ boxes. Referring to Appendix B 
for details, we find that there are only $n$ Young frames which contribute a non-vanishing character,
since non-vanishing contributions come from Young
frames with the property that all of the boxes lie on a single hook.
Furthermore, all of these have the value $\chi_{\lambda}(\sigma)=\pm 1$.
We therefore find
\beq 
I^{(N;N)}_n = n  \,\,\,\,\,\,\,\,\,\,\ ;~~n<N~,
\labell{longsimp2}
\eeq
giving the general expression
\beq
I^{(N;N)}_n = {\rm min}(N,n) \,\,\,\,\,\,\,\,\,\,\,\forall n,N~.
\labell{longtotal}
\eeq
 From this simple result, and the normalization factor ${\cal N}^{\prime(N;N)}=2/N$, we find
the amplitude
\beq
{\cal A}^{(N;N)} = \frac{2}{N}\sum_{n=0}^{\infty}(-\tilde{g})^{n}{\rm min}(N,n) 
= 2 f - {2\over N}\sum_{n=0}^{N-1}(N-n)(-\tilde{g})^n~.
\labell{longcoeff}
\eeq
in the case of closed string states of the form~\reef{ex1}. 

As another example let us consider states labelled by different conjugacy classes on 
the two sides of the string. Here we see an enormous simplification. For $n\ge N$ the orthogonality 
properties of characters of the symmetric group completely kills the integral, as indicated by 
the delta function in eqn.~\reef{sum1}. Thus
\beq
I^{(\sigma;\tilde{\sigma})}_{n} = 0 \,\,\,\,\,\,\,\,\, ; ~~ n\ge N~,
\labell{vanish}
\eeq
for $\sigma\neq{\tilde{\sigma}}$. Thus we have the remarkable result that these amplitudes 
only receive contributions from a finite number of terms in the perturbation series!

Let us determine these finite terms for $N=2$. The only amplitude
with different conjugacy class on each side is ${\cal A}^{(1^2;2)}$. Since
there are no contributions with $n\ge 2$ we simply need to evaluate
\beq
I^{(1^2;2)}_{n}=\sum_{\lambda\leq n}\chi_{\lambda}(\sigma)\chi_{\lambda}(\tilde{\sigma})~,
\labell{nlessm}
\eeq
for $n=1$. From the character table of $S_2$, 
\begin{table}
\begin{center}
\begin{tabular}{|c|c|c|} \hline
$\lambda \backslash\sigma$ & $(1^2) $& (2) \\ \hline
$(2) $ & 1 & 1 \\ \hline
$(1^2)$ & 1 & -1 \\ \hline
\end{tabular}
\caption{Character table for $S_2$. Each row gives $\chi_\lambda(\sigma)$ for a given $\lambda$.}
\label{s2table}
\end{center}
\end{table}
given explicitly in table (\ref{s2table}), we find
\beq
I^{(1^2;2)}_{1}=1~,
\labell{n1}
\eeq
and therefore the amplitude is
\beq
{\cal A}^{(1^2;2)}= {\cal N}^{\prime (1^2;2)}I^{(1^2;2)}=-\sqrt{2}~\tilde{g}~,
\labell{S2ans}
\eeq
since ${\cal N}^{\prime (1^2;2)}=\sqrt{2}$.

Proceeding similarly we have processed the characters of $S_3$ given in
table (\ref{s3table})
\begin{table}
\begin{center}
\begin{tabular}{|c|c|c|c|} \hline
$\lambda \backslash\sigma$ & (3) & $(2,1) $& $(1^3)$ \\ \hline
$(3) $ & 1 & 1 & 1 \\ \hline
$(2,1)$ & -1 & 0 & 2 \\ \hline
$(1^3)$ & 1 & -1 & 1 \\ \hline
\end{tabular}
\caption{Character table for $S_3$.}
\label{s3table}
\end{center}
\end{table}
and found all the closed string couplings up to level 3. The results are given in
table (\ref{ampltable}) where, for easy reference, we include also the
results from level 1,2.
\begin{table}
\begin{center}
\begin{tabular}{|c|c|} \hline
$(\sigma;\tilde{\sigma})$& ${\cal A}^{(\sigma;{\tilde\sigma})}$ \\ \hline
 $(1;1)$ & $2 [ f -1] $      \\ \hline 
$(2;2)$ &   $[2f-2+{\tilde g}]$      \\ \hline 
$(1^2,1^2)$ &$2[2f-2+{\tilde g}]$ \\ \hline
 $(1^2;2)$ & $\sqrt{2} [-{\tilde g}]$   \\ \hline
$(3;3)$ & ${2\over 3}[ 3f -3 + 2{\tilde g} - {\tilde g}^2] $      \\ \hline 
$(2,1;2,1)$ &   $2[2f-2+{\tilde g}-{\tilde g}^2]$      \\ \hline 
$(1^3,1^3)$ &${4\over 3}[6f-6+5{\tilde g}-{\tilde g}^2]$ \\ \hline
 $(2,1;1^3)$ & $\sqrt{8\over 3} [-{\tilde g}+{\tilde g}^2]$   \\ \hline
$(2,1;3)$ & $\sqrt{4\over 3} [-{\tilde g}+{\tilde g}^2]$   \\ \hline
$(3;1^3)$ & ${\sqrt{8}\over 3} [-{\tilde g}-{\tilde g}^2]$   \\ \hline
\end{tabular}
\end{center}
\caption{One point amplitudes of the closed strings up to level 3. The quantity in the square bracket is $I^{(\sigma;{\tilde\sigma})}$; the prefactor
is ${\cal N}^{\prime(\sigma;{\tilde\sigma})}$. The function $f=1/(1+{\tilde g})$.}
\label{ampltable}
\end{table}

\subsection{Comparison with Boundary States}

The boundary state for the rolling tachyon background takes the form
\beq
|B\rangle  = |B_{X^0}\rangle \otimes |B_{\stackrel{\rightarrow}{X}}\rangle \otimes |B_{ghost}\rangle~,
\labell{bstate}
\eeq
where $|B_{\stackrel{\rightarrow}{X}}\rangle $ is the usual boundary state for the spatial 
part of a D25 brane in bosonic string theory, $|B_{ghost}\rangle$ is the contribution from the ghosts
and
\beq
|B_{X^0}\rangle = B^{(0;0)}|0\rangle + B^{(1;1)}\al^0_{-1}\tilde{\al}^0_{-1}|0\rangle + 
{1\over\sqrt{2}}B^{(1,1;2)}(\al^0_{-1})^2\tilde{\al}^0_{-2}|0\rangle + \cdots~.
\labell{bstate2}
\eeq
The spatial and ghost components of the boundary state will play no role in this paper. The
temporal component of the boundary state can be computed following~\cite{cklm,RS,rolling} and gives
~\cite{lnt,GS}
\beq
|B_{X^0}\rangle = \sum_{j}  \sum_{m\geq 0} \left( 
\begin{array}{c} j+m\\2m
\end{array}  \right) (-{\tilde{g}})^{2m} | j,m,m \rangle\rangle~,
\eeq
where $| j,m,m \rangle\rangle $ are the Ishibashi states, {\it i.e.} infinite sets of states
built as Virasoro descendants of certain $SU(2)$ primaries. The laborious part of finding 
explicit expressions for the boundary states is to work out the Ishibashi states, since 
these become increasingly complex at higher level. In contrast our methods get at that information 
quite easily.

The boundary state coefficients $B^{(\sigma;\tilde{\sigma})}$ can be represented as the one 
point functions 
\beq
B^{(\sigma;\tilde{\sigma})} = \langle :V^{(\sigma;\tilde{\sigma})}:\rangle_{TBL}~,
\labell{bstate3}
\eeq
of the corresponding closed string vertex operators. It is the standard bulk normal 
ordering that appears in this expression, in contrast to the boundary normal
ordering which, as emphasized in~\cite{lnt}, appears in our amplitudes eq.\reef{finalans1}. The
two normal orderings are related by
\beq
:X^0(z,{\bar z})X^0(z^\prime,{\bar z}^\prime):~
=\no X^0(z,{\bar z})X^0(z^\prime,{\bar z}^\prime) \no + \log |z{\bar{z}}^\prime -1|~,
\label{normord}
\eeq
with the difference due to the "image" term in the disk 
Green's function eq.\reef{green2}. It follows, for example, that
\beq
: \partial^\nu X^0(0,0){\bar\partial}^{\nu^\prime} X^0(0,0):~
=\no \partial^\nu X^0(0,0){\bar\partial}^{\nu^\prime}X^0(0,0) \no 
-{1\over 2}\nu!(\nu-1)!\delta_{\nu\nu^\prime}~.
\eeq
Taking the normalization factors into account, and recalling that the
function $f$ is the vacuum amplitude, this gives
\beq
B^{(N;N)} = {\cal A}^{(N;N)} - f~,
\labell{bnn}
\eeq
or, from eq.\reef{longcoeff},
\beq
B^{(N;N)} = f  - {2\over N}\sum_{n=0}^{N-1}(N-n)(-\tilde{g})^n~.
\eeq
Let us emphasize that this expression gives the boundary state for any $N$.
The explicit computations of boundary states using Ishibashi states 
have been carried to level two~\cite{okuda}, with a result that agrees
with our general expression eq.\reef{bnn}. Extending the boundary computations 
to higher $N$ is however very tedious and in general not very practical. 

\begin{table}
\begin{center}
\begin{tabular}{|c|c|} \hline
$(\sigma;\tilde{\sigma})$& $B^{(\sigma;{\tilde\sigma})}$ \\ \hline
 $(1;1)$ & $ f - 2 $      \\ \hline 
$(2;2)$ &   $ f-2+{\tilde g} $      \\ \hline 
$(1^2,1^2)$ &$ f+2{\tilde g}$ \\ \hline
 $(1^2;2)$ & $-\sqrt{2}{\tilde g}$   \\ \hline
$(3;3)$ & $ f - 2  + {4\over 3}{\tilde g} - {2\over 3}{\tilde g}^2 $      \\ \hline 
$(2,1;2,1)$ &   $ f +  {\tilde g}-2{\tilde g}^2$      \\ \hline 
$(1^3,1^3)$ &$ f-2+{2\over 3}{\tilde g}-{4\over 3}{\tilde g}^2$ \\ \hline
 $(2,1;1^3)$ & $\sqrt{2\over 3} [{\tilde g}+2{\tilde g}^2]$   \\ \hline
$(2,1;3)$ & ${2\over \sqrt{3}} [-{\tilde g}+{\tilde g}^2]$   \\ \hline
$(3;1^3)$ & ${2\sqrt{2}\over 3} [-{\tilde g}-{\tilde g}^2]$   \\ \hline
\end{tabular}
\end{center}
\caption{Boundary state coefficients up to level 3.}
\label{bndytable}
\end{table}

The relation  eq.\reef{normord} between normal orderings can be extended to more 
complex operators. The general result has the same form as Wick's rule, 
except that contraction 
terms here appear only for operators with an identical number of derivatives. 
The two normal orderings are thus equivalent when $\sigma,\tilde{\sigma}$ have no 
cycles that are of the same length. 
The awkward numerical coefficient $-{1\over 2}\nu!(\nu-1)!\delta_{\nu\nu^\prime}$
that comes with each contraction conspires with the overall normalization
${\cal N}^{\prime(\sigma;{\tilde\sigma})}$ to give simple combinatorial
factors when relating $B^{(\sigma;\tilde{\sigma})}$
and ${\cal A}^{(\sigma;\tilde{\sigma})}$. For example, if only one type of
cycle appears
\beq
B^{(\nu^N;\nu^N)}={\cal A}^{(\nu^N;\nu^N)} -N{\cal A}^{(\nu^{N-1};\nu^{N-1})}
+ {N(N-1)\over 2}{\cal A}^{(\nu^{N-2};\nu^{N-2})}+\cdots~,
\eeq

We have used the contraction rules to compute boundary state coefficients from our 
matrix amplitudes for all states up to level three. The results are given in table (\ref{bndytable}).
As a check we have carried the boundary state computations one level higher
than~\cite{okuda}, to level three, and verified agreement of all terms. This
gives us great confidence that the two methods really are equivalent. It
also shows that the matrix method is by far the most convenient.

%%%%%%%%%%%%%%%%%%%%%%%%%%%%%%%%%%%%%%%%%%%%%%%%%%%%%%%%%%%%%%%%%%%%%%%%%%%%%%%%%%%%%%%%%%%%%%
\section{Bulk Correlators}

In this section we consider amplitudes of the form
\beq
{\cal A}^{\Pi\exp(-n_kX^0)} = \langle : \prod_{k=1}^{m}e^{-n_kX^0(z_k,\bar{z}_k)}: \rangle_{TBL}~,
\eeq
where the $z_k$ are in the interior of the unit disk.
These amplitudes are the building blocks of general bulk correlators in the 
background of the rolling tachyon. They were previously considered in~\cite{GS} where explicit results 
were presented for $m\le 2$. We present these here both to reiterate our general theme that 
all correlation functions in this theory are matrix integrals and to demonstrate the ease with which 
the resulting integrals can be evaluated, even in the general case.

Although it is not strictly necessary for our methods to work, we will enforce momentum conservation.
Then the only contribution to the amplitude is at $n$th order in perturbation theory where $n=\sum n_k$; 
and so, up to an overall factor of $(-2\pi g)^n$, the entire amplitude reduces to 
\beq
{\cal A}_{n}^{\Pi\exp(-n_kX^0)} = {1\over n!}\langle\prod_{k=1}^{m}e^{-n_kX^0(z_k,\bar{z}_k)}
\prod_{i=1}^{n}\int\frac{dt_i}{2\pi}e^{X^0(w_i,\bar{w}_i)}\rangle~,
\labell{bulkcorr}
\eeq
where the $w_i$ are situated on the boundary.
Straightforward calculation of the contractions, using the Green's function 
eq.\reef{green2}, gives
\begin{eqnarray}
{\cal A}_{n}^{\Pi\exp(-n_kX^0)} &=& {1\over n!}\prod_{k=1}^{m}|z_k\bar{z}_k-1|^{n_k^2/2}\prod_{k<l}|z_k-z_l|^{n_kn_l}
|z_k\bar{z}_l-1|^{n_kn_l} \times ~ \\ \nonumber
&~&~~~~~~~~\times \prod_{i=1}^{n}\int {dt_i\over 2\pi}
\prod_{i<j}|e^{it_i}-e^{it_j}|^2\prod_{k=1}^m\prod_{i=1}^n|z_k-w_i|^{-2n_k}~,
\labell{bulkamp}
\end{eqnarray}
where we have used $\bar{w}=1/w=e^{-it}$ on the boundary of the unit disk. The last term in the 
integrand can be rewritten
\beq
\prod_{k=1}^m\prod_{i=1}^n |z_k-w_i|^{-2n_k}=
\prod_{k,i}(1-z_ky_i)^{-n_k}(1-\bar{z}_k\bar{y}_i)^{-n_k}~,
\labell{symmfunc}
\eeq
where we have defined $y_i = 1/w_i$ and used the fact that $w_i\bar{w}_i=1$.
But the product
\beq
h(z_k)\equiv \prod_{i}(1-z_ky_i)^{-1} = {\rm det}(1-z_kU^{\dagger})^{-1} ~, 
\labell{symm}
\eeq
is expressed directly in terms of matrices; and so
\beq
{\cal A}_{n}^{\Pi\exp(-n_kX^0)} = \prod_{k=1}^{m}|z_k\bar{z}_k-1|^{n_k^2/2}\prod_{k<l}|z_k-z_l|^{n_kn_l}
|z_k\bar{z}_l-1|^{n_kn_l}\int \frac{dU}{{\rm vol}(U(n))}\prod_k h(z_k)^{n_k}h(z_k)^{\dagger n_k}~,
\labell{simple}
\eeq
which is the advertized formula for this entire class of correlators, written in terms of unitary matrix 
integrals. 

 In order to evaluate the integral eq.\reef{simple} first recast $h(z_k)^{n_k}$
in a more convenient form by defining a new set of variables
\beq
\{Z_K\} = (\overbrace{z_1,z_1\cdots,z_1}^{n_1};\overbrace{z_2,z_2,\cdots,z_2}^{n_2}
\cdots;\overbrace{z_m,z_m,\cdots,z_m}^{n_m})~,
\labell{stretch}
\eeq
where the index $K\in (1,n)$ with $n=\sum_{i}n_i$. This enables us to write
\beq
\prod_{k=1}^{m}h(z_k)^{n_k} = \prod_{k,i}(1-z_ky_i)^{-n_k} = \prod_{K,i}(1-Z_{K}y_i)^{-1}=
\prod_{K} h(Z_K)~.
\labell{stretchsym}
\eeq
Now, the expression eq.\reef{symm} is in fact the textbook definition of the generating 
function for what are known as the complete symmetric polynomials -- see Appendix C. 
These are fundamental objects in the theory of  symmetric functions since, among other things, 
they provide a basis for the ring of symmetric polynomials. Also of central importance is
the Cauchy identity
\beq
\prod_{K,i}(1-Z_{K}y_i)^{-1} = 
\sum_{\lambda}s_{\lambda}(Z)s_{\lambda}(y)~,
\labell{cauchy}
\eeq
where $s_{\lambda}(x)$ are the Schur functions for the abstract variables $x=\{x_i\}$, 
labelled by partitions (Young frames) $\lambda$. In the case we are considering 
$y_i=e^{-it_i}$ are the eigenvalues of the matrix $U^{\dagger}$ and therefore
the Schur function $s_{\lambda}(y)$ is known to be equivalent to the character of $U^{\dagger}$ in the irreducible representation
labelled by the partition $\lambda$ {\it i.e.,} $s_{\lambda}(w) = \chi_{\lambda}(U^{\dagger})$. We will not 
need the explicit form of the Schur functions for the variable $Z =\{Z_{K}\}$. We may now write
\beqa
\frac{1}{{\rm vol}(U(n))}\int dU\prod_{K} h(Z_K)h(Z_K)^{\dagger}  
&=& \sum_{\lambda\lambda^{\prime}}
s_{\lambda}(Z)s_{\lambda^{\prime}}(\bar{Z})
 \frac{1}{{\rm vol}(U(n))}\int dU \chi_{\lambda}(U)\bar{\chi}_{\lambda^{\prime}}(U)  \nonumber \\
&=&\sum_{{\lambda\lambda^{\prime}}}s_{\lambda}(Z)s_{\lambda^{\prime}}(\bar{Z})
\delta_{\lambda\lambda^{\prime}}\nonumber \\
&=& \sum_{{\lambda}}s_{\lambda}(Z)s_{\lambda}(\bar{Z})~,
\labell{orthcauchy}
\eeqa
where we have once again evaluated the integral over $U(n)$ using the orthogonality 
of group characters. Instead of computing the remaining sum term by term
we can use the Cauchy identity in reverse to obtain
\beq
\sum_{{\lambda}}s_{\lambda}(Z)s_{\lambda}(\bar{Z})= 
\prod_{I,J=1}^{n}(1-Z_{I}\bar{Z}_{J})^{-1}
=\prod_{i,j=1}^m(1-z_i\bar{z_j})^{-n_in_j}~.
\labell{reverse}
\eeq
Assembling eqs.~\reef{simple},\reef{orthcauchy} and~\reef{reverse} we obtain the general form of the correlator
\beq
{\cal A}_{n}^{\Pi\exp(-n_kX^0)} = \prod_{k=1}^{m}|z_k\bar{z}_k-1|^{-n_k^2/2}\prod_{i<j}^m|z_i-z_j|^{n_in_j}
\prod_{i<j}^{m}\left|z_i\bar{z}_{j}-1\right|^{-n_in_j}~. 
\labell{bulkfinal}
\eeq
This generalizes the result of~\cite{GS} to include all $m$-point functions of bulk tachyons for $m>2$. 
Note that we have made no recourse here to the $SU(2)$ current algebra, nor to contour integration.

\section{Boundary Correlators}

For completeness we point out that correlators involving insertions of vertex operators on the
boundary of the disk can also be written as matrix integrals. In fact this follows almost trivially
from the previous section. The general boundary correlator is
\beq
\tilde{{\cal A}}_{n}^{\Pi\exp(-n_kX^0)} = \frac{1}{n!}\langle\prod_{k=1}^{m} \no e^{-n_kX^0(z_k,\bar{z}_k)}\no
\prod_{i=1}^{n}\int\frac{dt_i}{2\pi}e^{X^0(w_i,\bar{w}_i)}\rangle~,
\labell{boundcorr}
\eeq
where now the $z_k$ are points on the boundary of the disk. The contractions
then give
\beq
\tilde{{\cal A}}_{n}^{\Pi\exp(-n_kX^0)} =\frac{1}{n!}\prod_{k<l}^{m}|z_k-z_l|^{2n_kn_l}\prod_{i=1}^{n}\int {dt_i\over 2\pi}
\prod_{i<j}|e^{it_i}-e^{it_j}|^2\prod_{k=1}^m\prod_{i=1}^n|z_k-w_i|^{-2n_k}~.
\labell{boundamp}
\eeq 
The simplifications relative to eq.~\reef{bulkamp} are due to $|z|=1$. Another difference
is that we now use boundary normal ordering to regulate the vertex operators. 
We find
\beq
\tilde{{\cal A}}_{n}^{\Pi\exp(-n_kX^0)} =\prod_{k<l}^{m}|z_k-z_l|^{2n_ln_k}\int \frac{dU}{{\rm vol}(U(n))}
\prod_k h(z_k)^{n_k}h(z_k)^{\dagger n_k}~,
\labell{simple2}
\eeq
with the same definitions as in section five above. Proceeding as before the 
boundary amplitude becomes
\beq
\tilde{{\cal A}}_{n}^{\Pi\exp(-n_kX^0)}  = 
\prod_{k=1}^{m}|z_k\bar{z}_k-1|^{-n_k^2}~.
\labell{boundfinal}
\eeq
This expression is clearly divergent since $z_k\bar{z}_k = 1$ for points on the boundary. 
Dealing with this divergence directly 
in timelike boundary Liouville theory is difficult and is perhaps best dealt with by 
reverting to analytic continuation from the spacelike boundary Liouville theory~\cite{fzz} as advocated in~\cite{GS}.
We are hopeful however that the matrix integral perspective may circumvent this indirect approach.

%%%%%%%%%%%%%%%%%%%%%%%%%%%%%%%%%%%%%%%%%%%%%%%%%%%%%%%%%%%%%%%%%%%%%%%%%%%%%%%%%%%%%%%%

\section{Rolling Tachyons and $QCD_2$}

In section three we observed that the one point functions of massive string 
states in the background of the
rolling tachyon take the form,
\beq
\langle \no V^{(\sigma;\tilde{\sigma})}\no\rangle_{TBL} = {\cal N}^{\prime(\sigma;\tilde{\sigma})}
\sum_{n=0}^{\infty}(-\tilde{g})^nI_n^{(\sigma;\tilde{\sigma})}~,
\labell{onepoint}
\eeq
where the summand is given by the $U(n)$ integral,
\beq
I_n^{(\sigma;\tilde{\sigma})} = \frac{1}{{\rm vol}(U(n))}\int dU \prod_{i,\tilde{i}}[Tr(U^{\nu_i})]^{N_i}
[Tr(U^{\dagger\tilde{\nu}_{\tilde{i}}})]^{\tilde{N}_{\tilde{i}}}~.
\labell{wilson}
\eeq
As we already noted, this is {\it exactly} the correlation function of a pair of Wilson loops in
two dimensional $U(n)$ gauge theory in the limit where the area of the
two dimensional manifold vanishes~\cite{wati1,wati2}.
The precise correspondence is somewhat unusual since from eqn.~\reef{onepoint} the full
correlator in the rolling tachyon background is given by the sum over Wilson loop 
correlators in different gauge theories! In other words,
\beq
\langle \no V^{(\sigma;\tilde{\sigma})}\no\rangle_{TBL} = 
\sum_{n=0}^{\infty}(-\tilde{g})^n
\langle W_n(\sigma)\bar{W}_n(\tilde{\sigma})\rangle_{QCD_2}~.
\labell{correspond}
\eeq
where $W_n(\sigma)$ is a Wilson loop operator in two dimensional $U(n)$ gauge theory.

Since correlators involving other vertex operators in the rolling tachyon background  
similarly reduce to matrix integrals we make the following proposal:
given a set of vertex operators $V_i(z,\bar{z})$ then general correlation functions in the
timelike boundary Liouville theory can be expressed as
\beq
\langle  \prod_{i}V_i(z,\bar{z})\rangle_{TBL} = \sum_{n=0}^{\infty}(-\tilde{g})^n
\langle \prod_{i}Q^n_{i}(z;\alpha)\bar{Q}^n_{i}(\bar{z};\tilde{\alpha})\rangle_{QCD_2}~,
\labell{guess}
\eeq
where $Q^n_{i}(z;\alpha)$ is a gauge invariant operator in $U(n)$ 
Yang-Mills with $\alpha$ representing 
how this operator depends on the details of the Liouville 
vertex operators $V_i(z,\bar{z})$.
As an empirical formula~\reef{guess} is certainly valid 
in the cases we have considered in this paper. 
The examples we have uncovered thus far
can be usefully organized as
\beqa
\prod_{i}\left[\prt^{\nu_i}X^0(0)\right]^{N_i}&\leftrightarrow&
\prod_{i}\left[\Tr(U^{\dagger\nu_i})\right]^{N_i} 
\leftrightarrow \sum_{\lambda}\chi_{\lambda}(\sigma)\chi_{\lambda}(U^{\dagger})~,\nonumber \\
\exp(-X^0(z)) &\leftrightarrow& \det(1-zU^{\dagger})^{-1} \leftrightarrow 
\sum_{\lambda}s_{\lambda}(z)\chi_{\lambda}(U^{\dagger})~,
\labell{maps}
\eeqa
where we have focused on the holomorphic parts of the vertex operators in question 
and the implied correspondence holds at the level of correlation functions. 
Viewed this way it is clear that the characters of the unitary group $\chi_{\lambda}(U)$ are playing the
role of a complete set of functions into which all the operators can be expanded. The expansion
coefficients on the other hand are dependent on the specific vertex operator. It would be
interesting to uncover the
general rule for associating Liouville vertex operators 
with gauge invariant operators of $QCD_2$. 

It is perhaps useful to explain in more detail how our envisioned correspondence
is realized. First, we recall the interpretation of $QCD_2$ as a string theory~\cite{wati1}.
The fundamental idea is that the partition function can be understood as  
the sum over all maps of a two dimensional world sheet onto a two
dimensional target space of fixed
topology. The maps in question are usually referred to as covering maps and have an associated
winding number $N$ indicating that the map covers the target space $N$ times. 
The physical observables of any gauge theory are the gauge invariant 
operators, a particular example of which is the Wilson loop,
\beq
W(\sigma) = \Tr\left(e^{\oint_{\sigma} A\cdot dx}\right) \equiv \Tr(U_{\sigma})   .
\labell{wilsonloop}
\eeq
where the matrix $U_{\sigma}$ represents the holonomy of the gauge field as one traverses
the path defined by $\sigma$. 
In the stringy realization of $QCD_2$ a Wilson loop is an $S^1$ inside of the
target space onto which the boundary of
a worldsheet is mapped. Since the target space will in general 
have a finite number of punctures the map 
will have branch points. As one traverses the $S^1$ of the target
space one will encounter branch points where the different sheets of the map meet. The
path is thus determined by specifying how the sheets are permuted into each other. 
Therefore, in addition to the winding number $N$, we also label each Wilson loop by 
an element $\sigma\in S_N$. Here $\sigma$ should be thought of as defining the path by specifying 
how the sheets are permuted as one
traverses the Wilson loop. In string theory language the natural set of
gauge invariant observables are~\cite{wati2},
\beq
\Tr(U_{\sigma}) = \prod_{i}\left[\Tr(U^{\nu_i})\right]^{N_i} 
\labell{stringyloop}
\eeq
where on the right hand side $U$ is taken to be in the defining representation
of $U(n)$ and $(\nu_i;N_i)$ specify a partition of $N$.
With this suggestive notation it is clear how we should translate 
between the $QCD_2$ language and 
that of the  rolling tachyon: we identify, respectively, $N$ 
and $\sigma$ with the level number and oscillator structure of the 
massive closed string whose one point function we are calculating 
in the rolling tachyon background. 
The interpretation is that these one point functions 
of timelike boundary Liouville theory are encoding information about how 
Riemann surfaces with boundaries can be mapped into Riemann 
surfaces with a fixed number of punctures.  
Note that the correspondence we are suggesting 
is strictly valid only in the limit of vanishing area of the target space. 
Thus, intuitively, the information contained in the Liouville theory should only be
topological. Also, since we must include all values of $n$ it is not completely clear
how the geometric interpretation should manifest itself since the above stringy interpretation of 
$QCD_2$ is usually associated with a large $n$ limit. We leave investigation of these interesting issues
to future work.

%%%%%%%%%%%%%%%%%%%%%%%%%%%%%%%%%%%%%%%%%%%%%%%%%%%%%%%%%%%%%%%%%%%%%%%%%%%%%%%%%%%

\section*{Acknowledgements}
NRC would like to thank Washington Taylor, Jan Troost and especially 
Alexander Postnikov for useful discussions. NRC would also like to 
thank the Michigan Center for Theoretical Physics for its hospitality
during the initial stages of this project. FL thanks Asad Naqvi and
Seiji Terashima for the collaboration on~\cite{lnt} which raised the ideas
developed here. NRC is supported by the DOE under grant DF-FC02-94ER40818, the NSF under
grant PHY-0096515 and NSERC of Canada. FL was supported in part by the DoE.

%%%%%%%%%%%%%%%%%%%%%%%%%%%%%%%%%%%%%%%%%%%%%%%%%%%%%%%%%%%%%%%%%%%%%%%%%%%%%%%%%%%%%%%%%%%

\section*{Appendix A: Schur-Weyl Duality}

The purpose of this appendix is to provide an introduction to Schur-Weyl duality,
used repeatedly in the main text. For more extensive discussion and formal proofs
we refer to any one of a large number of standard texts on the subject.
The discussion presented here largely follows ref.~\cite{schurweyl}. 

The basic observation of Schur and Weyl is that there is a deep connection between the
representation theory of the symmetric group $S_N$ and that of the unitary group $U(n)$. 
The defining representation of $U(n)$ is given by $n\times n$ unitary 
matrices which act on an $n$-dimensional
complex vector space $V$ which is referred to as the carrier 
space of the representation. One now considers the
$N$-fold tensor product space ${\cal V}_N= V\otimes V\otimes\cdots\otimes V$ 
which is acted on by the group 
$H=U(n)\times S_N$. Here $S_N$ is the symmetric group of $N$ 
objects and acts by permuting the $N$ factors
appearing in ${\cal V}_N$. Given an arbitary element
 $h=\sigma\times U \in H$ where $\sigma\in S_N$ and 
$U\in U(n)$ then $h$ in its defining representation $D(h)$
  acts on the basis vectors $v_{i_k}$ 
in the product space as
\beq
v_{i_1}v_{i_2}\cdots v_{i_N} \stackrel{D(h)}{\rightarrow}
 U_{i_1\sigma(j_1)}U_{i_2\sigma(j_2)}\cdots 
U_{i_N\sigma(j_N)}v_{j_1}v_{j_2}\cdots v_{j_N} \equiv 
U_{(i)(j)}^{\sigma}v_{j_1}v_{j_2}\cdots v_{j_N}~,   
\labell{basisaction}
\eeq
where the notation $(i)\equiv (i_1,i_2\cdots i_N)$ has been
 used and the superscript $\sigma$ indicates
that the permutation is to be implemented on the set of integers $(j)$. 
Clearly the actions of the abstract group elements $\sigma$ and $U$ commute 
with each other. From the point of view of the group $U(n)$  
one would like to understand what are the invariant
subspaces. Put differently, the action of the unitary group on the
 tensor product space is in general reducible
and one would like to know which irreducible representations of $U(n)$ appear in the decomposition.
Schur-Weyl duality answers this question. 

To see how this comes about note that the irredicble representations 
of $S_N$ correspond  to subspaces of ${\cal V}_N$ that are invariant (under $S_N$) . 
Now, the irreducible representations of $S_N$ are in a one to one 
correspondence with the partitions of $N$ into integers which, in turn,
are nicely represented in terms of Young frames. Schur-Weyl duality 
is the statement that the $U(n)$ invariant subspaces of ${\cal V}_N$ are in a one to 
one correspondence with the $S_N$ invariant subspaces. Operationally this means 
the irreducible representations of $U(n)$ appearing in the decomposition are precisely those
corresponding to the Young frames classifying the irreducible representations of $S_N$,
with each $U(n)$ representation appearing precisely once.

It is perhaps instructive to see this in a simple example~\cite{schurweyl}. 
Consider the case $N=2$ so that the  elements of ${\cal V}_2$ are given
by the second rank tensor $F_{ij}$. If we represent the elements of $S_2$ by
$(e,s)$ then we have $eF_{ij} = F_{ij}$ and $sF_{ij} = F_{ji}$. The irreducible 
representations of $S_2$ are summarized  by the Young frames depicted in figure
~\reef{figs2}.
\begin{figure}[ht!]
\center{\includegraphics{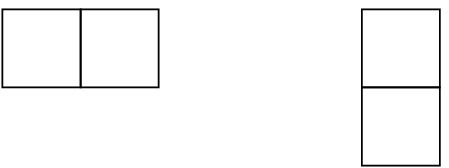}}
\caption{Young frames for $S_2$.} 
\labell{figs2}
\end{figure}
Now, the operators $e+s$ and $e-s$ (known as Young operators) act on the $F_{ij}$ as
projections on to the symmetric and antisymmetric pieces of $F_{ij}$, respectively, which are 
clearly the $S_2$ invariant subspaces of ${\cal V}_2$. This is also clearly what one expects 
as the invariant subspaces under $U(n)$. Since both $e+s$ and $e-s$ commute with the 
action of $U(n)\times U(n)$ they also project out invariant subspaces of the carrier space 
${\cal V}_2$. The irreducible representations of $U(n)$ which occur are then also completely 
summarized in figure~\reef{figs2}. 

For general $S_N$ we can similarly think of the defining representation $D(h)$ of 
$h=\sigma\times U$ as a tensor representation which in general is reducible and 
decomposes according to the irreducible representations of $S_N$ as
\beq
D(h) = \sum_{\lambda}S_{\lambda}(\sigma)\otimes T_{\lambda}(U)~,
\labell{decomp}
\eeq
where $S_{\lambda}(\sigma)$, $T_{\lambda}(U)$ are the irreducible representations 
of $\sigma$ and $U$ appearing
in the decomposition. Again the index $\lambda$ should be thought of as indexing the 
Young frames defining the irreducible representations of $S_N$. It is clear from this tensorial
point of view that the entries on the right hand side of eq.\reef{decomp} corresponding
to diagrams with columns of height more than $n$ actually vanish as representations
of $U(n)$. Note that we keep columns with height exactly $n$ because we are considering 
$U(n)$, rather than $SU(n)$. 
The reasoning appealed to here is more or less the standard one leading to the use
of Young frames to label $U(n)$ representation. Of course a more rigourous treatment
is possible, see~\cite{weyl}. 

Our main interest is the consequences of eq.\reef{decomp} for the characters.
Taking the trace, we find
\beq
\Tr(D(h)) = \sum_{\lambda}\chi_{\lambda}(\sigma)\chi_{\lambda}(U)~,
\label{chardecomp}
\eeq
since the character of a direct product is the product of the characters. Note that,
since the character is the same for all elements of a conjugacy class, we can
now think of the $\sigma$ in the argument of $\chi_\lambda$ as a conjugace class,
rather than a group element. The left hand side can be more explicit by considering 
the expression eq.~\reef{basisaction} for the defining representation $D(h)$ and 
directly taking the trace of the operator $U_{(i)(j)}^{\sigma}$, by setting $(i)=(j)$ 
and summing over $(j)$ . We find
\beq
\Tr U_{(i)(j)}^{\sigma} = \sum_{(j)}U_{(j)(j)}^{\sigma} = \prod_k[\Tr(U^{\nu_k})]^{N_k}~,
\labell{idenity}
\eeq
where the permutation $\sigma$ contains $N_1$ cycles of length $\nu_1$, $N_2$ cycles of length $\nu_2$ etc.
For example if $\sigma=(1^N)$ {\it i.e.,} the identity permutation $\sigma=e$, then clearly
\beq
\Tr U_{(i)(j)}^{(1^N)}= \sum_{j_1}\sum_{j_2}\cdots\sum_{j_N}U_{j_1j_1}U_{j_2j_2}\cdots U_{j_Nj_N} = [\Tr U]^N~,
\labell{exam1}
\eeq
and likewise, if $\sigma=(N)$, then 
\beq
\Tr U_{(i)(j)}^{(N)}= \sum_{j_1}\sum_{j_2}\cdots\sum_{j_N}U_{j_1j_2}U_{j_2j_3}\cdots U_{j_Nj_1} = 
\Tr(U^N)~.
\labell{exam2}
\eeq
Generally, we find
\beq
\prod_k[\Tr(U^{\nu_k})]^{N_k}=\sum_{\lambda}\chi_{\lambda}(\sigma)\chi_{\lambda}(U)~,
\labell{identity}
\eeq
which is the formula used extensively in the main body of the paper, to simplify integrals .

\section*{Appendix B: The Murnaghan-Nakayama Rule.}
Our algebraic algorithm for determining the couplings to closed strings ultimately
relies on the computation of characters of the symmetric group. In this section
we review a simple graphical technique for doing this, known as the Murnaghan-Nakayama 
rule. (For a derivation see~\cite{symfunc}.)
With this rule in hand the industrious reader may verify the explicit results quoted in
the main text, or find more general ones. 
 
The goal is to calculate the characters $\chi_{\lambda}(\sigma) = \Tr_{\lambda}(\sigma)$. We first
 recall two fundamental facts about the 
symmetric group. First, the conjugacy classes $\sigma$ of $S_N$, which are defined 
by specifying a cycle structure,  are in a one to one correspondence with the partitions of 
$N$. Second, the irreducible representations $\lambda$ of $S_N$ are also in a one to 
one correspondence with the partitions of $N$. Thus the input required to calculate the 
character of a given conjugacy class in a given representation is simply a pair 
of partitions of $N$. 
In the following we will use two techniques to encode a partition of $N$. 
First, we will identify a given representation
$\lambda$ with a Young frame as follows. We can always write a partition as 
$N = \lambda_1 +\lambda_2 +\cdots \lambda_k$ where
by convention we take $\lambda_1\ge\lambda_2\ge\cdots\ge\lambda_k$. 
We then construct the Young frame by drawing a row of boxes
of length $\lambda_1$. Immediately under this (and aligned on the left)
 we draw a row of boxes of length $\lambda_2$. We continue this
until we have a row of boxes for each of the $\lambda_i$. Each frame will 
(by definition) contain $N$ boxes. Each 
possible shape corresponds to a different partition of $N$ and thus a specific irreducible
representation $\lambda$. Next, a given conjugacy class $\sigma$ is specified
 by its cycle structure and we can therefore
label a conjugacy class by specifying the lengths of it's cycles {\it i.e.,}
 $\sigma = (\mu_1,\mu_2,\cdots\mu_k)$ where
$\sum_{i}\mu_i=N$. 
Note that the notation used in this rule differ from $\sigma=(\nu_1^{N_1},\cdots,\nu_k^{N_k})$
used previously by introducing a distinct label for each of several cycles having the
same length.
With this input the character of $\sigma$ in the
 representation $\lambda$ can be calculated with the
following rule:

\begin{itemize}
\item 
Draw the Young frame correponding to $\lambda$. 
\item
Fill in the boxes with $\mu_1$ $1$'s, $\mu_2 $ $2$'s, $\mu_3$ $3's$, $\mu_4$ $4$'s etc. so that
\begin{enumerate}
\item Each set of numbers forms a continous hook pattern. By this we mean
an  uninterrupted vertical line, followed by a horizontal line to the right;
or a horizontal line to the right, follwed by a vertical line up. 
\item The numbers are weakly increasing from left to right and top to bottom.
\end{enumerate}
\item
Each table is assigned the number $H=(-1)^{\sum(h_i-1)}$ where $h_i$ is the height of each hook,
{\it i.e.} the number of boxes in the vertical part of the hook.
\item
Repeat the above procedure for all possible ways of covering the Young frame with hooks according to
the above rules.
\item
The character $\chi_{\lambda}(\sigma)$ is given by the sum of all the $H$ values for each covering. 
\end{itemize}

The computations using the Murnaghan-Nakayama rule are in fact simpler than it may appear
at first. Let us consider an explicit example. Take the representation $\lambda=(4,3,2)$ of
$S_9$ and let us evaluate its character on the group element $\sigma = (12)(345)(678)(9)$ 
which belongs to the conjugacy class which we can write  as $\sigma=(2,3,3,1)$, where the entries indicate 
the cycle lengths. Notice $2+3+3+1 = 9$ and also $4+3+2=9$; so these are partitions of $9$. 
The procedure for computing the character $\chi_{(4,3,2)}(2,3,3,1)$ is then spelled out 
in figure~\reef{figMN}. Each table has the shape corresponding
to the representation $(4,3,2)$ and there are only two possible coverings with numbers
consistent with the rules given above. Each of these tables has $H=1$; so
$\chi_{(4,3,2)}(2,3,3,1)=2$. 

\begin{figure}[ht!]
\center{\includegraphics{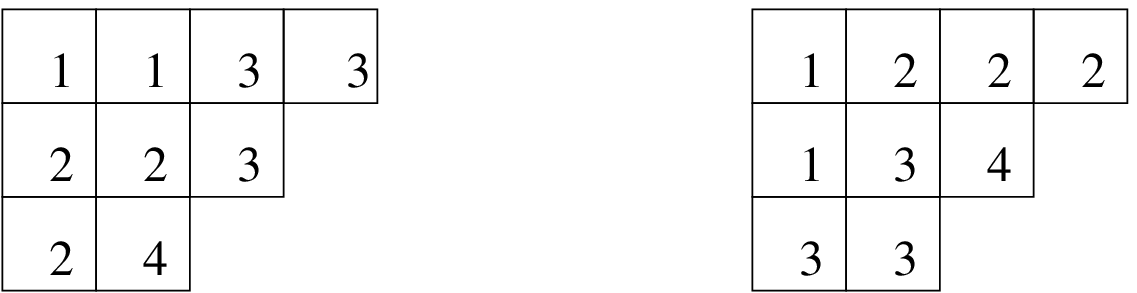}}
\caption{A demonstration of the Murnaghan-Nakayama Rule.} 
\label{figMN}
\end{figure}

\begin{figure}[ht!]
\center{\includegraphics{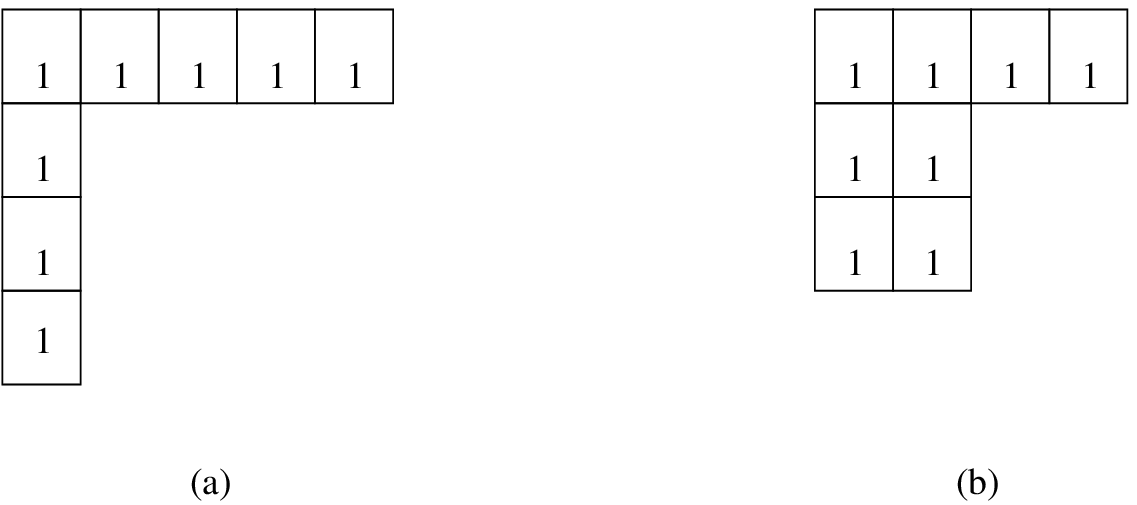}}
\caption{(a) A frame which allows a continuous hook, and thus gives a non-vanishing 
character.
(b) A frame which does not allow any continuous hooks, and thus 
gives a vanishing character.}
\label{figY2}
\end{figure}

As another class of examples, consider the conjugacy class of long cycles in $S_N$ {\it i.e.,}
$\sigma=(N)$ in our present notation. There are lot's of Young frames with $N$
boxes; but we need to cover a given frame with a hook of $N$ $1$'s and this leaves 
only single hook diagrams. For example
figure (\ref{figY2}a) can accomodate a hook, but figure (\ref{figY2}b) cannot. For each 
frame allowing a hook there is obviously only one way to cover the frame so, whatever the 
height, $H= (-)^{h_1-1}=\pm 1$. Thus, as claimed in the main text, $\chi_{\lambda}(N)=\pm 1$ 
for all the representations with non-zero character on the long cycle. 
If we restrict to representations with no more
than $n$ boxes in a column there are precisely $n$ such frames and so 
$\sum_{\lambda\le n} \chi_{\lambda}^2(N)=n$.

Explicit computations using the Murnaghan-Nakayama rule can obviously become
quite laborious for complex representations. The point we want to make here is simply
that the computation of amplitudes is completely algebraic. Indeed, for practical 
computations at high level there even exists a MAPLE code~\cite{maple} which 
automates the determination of the characters, and thus the amplitudes.
%\begin{itemize}
%\item Exercise: Derive the character table for $S_3$ using the Murnaghan-Nakayama rule.
%\end{itemize}

%%%%%%%%%%%%%%%%%%%%%%%%%%%%%%%%%%%%%%%%%%%%%%%%%%%%%%%%%%%%%%%%%%%%%%%%%%%%%%%%%%%%%%%%%%%%
\section*{Appendix C: Schur Functions and the Cauchy Identity}

This appendix will attempt to summarize some basic facts and definitions from the theory of 
symmetric functions that are useful for evaluating various integrals in the main sections of the
paper. For proofs and derivations there is a clear discussion of this enormous subject in ref.~\cite{symfunc}.

Consider a partition $\lambda$ of the form $\lambda_1 + \cdots + \lambda_n=N$ where,
in the present context, some of the $\lambda_i$ may vanish.
We can then define a symmetric polynomial in 
$n$ variables and degree $N$ as 
\beq
m_{\lambda} =  \sum_{\sigma\in S_n}x_1^{\sigma(\lambda_1)}x_2^{\sigma(\lambda_2)}\cdots x_n^{\sigma(\lambda_n)}~.
\labell{monomial}
\eeq
For $N\leq n$ the special case $\lambda_i = 1$ ($i\leq N$)  and $\lambda_i=0$ ($N<i\leq n$) 
gives rise to what are known as the elementary symmetric functions in $n$ variables
\beq
e_{N} = \sum_{1\le i_1 < i_2\cdots<i_N\le n}x_{i_1}x_{i_2}\cdots x_{i_N}~,
\labell{elem}
\eeq
and can be represented by the generating series
\beq
e(t) = \sum_{N\ge 0} e_N t^N = \prod_{i=1}^{n}(1+tx_i)~.
\labell{elgen}
\eeq
Another class of symmetric functions are the complete symmetric functions $h_k$. They 
are defined as
\beq
h_N=\sum_{1\le i_1 \le i_2\cdots\le i_N\le n}x_{i_1}x_{i_2}\cdots x_{i_N}~,
\labell{comp}
\eeq
and they have the generating function
\beq
h(t) = \sum_{N\ge 0}h_Nt^N = \prod_{i=1}^{n}(1-tx_i)^{-1}~.
\labell{compgen}
\eeq
Note that $e(t)h(-t)=1$.
 
It is also useful to consider the anti-symmetric functions defined by
\beq
a_{\lambda}=\sum_{\sigma\in S_n}\epsilon(\sigma)x_1^{\sigma(\lambda_1)}x_2^{\sigma(\lambda_2)}\cdots x_n^{\sigma(\lambda_n)}~,
\labell{anti}
\eeq
where $\epsilon(\sigma)$ is the sign of the permutation $\sigma$. These functions are completely 
antisymmetric under the interchange of any two of the variables $x_i$.  The sets of 
symmetric and anti-symmetric functions are isomorphic to each other. The isomorphism can
be realized concretely as multiplication by the 
Vandermonde determinant
\beq
\Delta(x) = \det(x_i^{n-j}) = \prod_{1=i<j}^{n}(x_i-x_j)~,
\labell{vanderapp}
\eeq
where the notation $x_i^{n-j}$ is understood to indicate the $(i,j)$ entry of an $n\times n$ matrix.

With the above definitions we now come to the Schur functions. These are a very general class of symmetric functions which 
contain as special cases both the elementary and complete symmetric functions above. A Schur function $s_{\lambda}(x)$ 
for some abstract variables $x_i$ is specified by a partition $\lambda$ (of the same form as above) as 
\beq
s_{\lambda}(x) = \frac{\det(x_i^{\lambda_j + n-j})}{\det(x_i^{n-j})}~.
\labell{schurdef}
\eeq
The Schur functions are symmetric polynomials of degree $N$.  
For example, if $\lambda_1 = N$ and $\lambda_i = 0, i\in (2,n)$ then $s_{\lambda}(x) $ reduces to the 
complete symmetric function $h_{N}$. When $\lambda_i = 1,i\in (1,n)$ \reef{schurdef} gives the elementary 
symmetric fucntion $e_N$. For the purposes of
this paper the Schur functions are useful because of their connection to 
the characters of the unitary group. When the variables $x_i = e^{it_i}$ 
it turns out that $s_{\lambda}(x) = \chi_{\lambda}(U)$, where $U$ is the unitary matrix for 
which the $e^{it_i}$ are the eigenvalues. 

We now state, without proof, the Cauchy identity. This is a remarkable identity which relates {\it products}
of complete symmetric functions to Schur polynomials. Consider a set of generating functions for complete 
symmetric functions $h(z_l),\,\,l\in (1,m)$. Then the product over this set can be written in terms of Schur functions as
\beq
\prod_{l=1}^mh(z_l) = \prod_{l=1}^{m}\prod_{i=1}^{n}(1-z_lx_i)^{-1} = \sum_{\lambda}s_{\lambda}(z)s_{\lambda}(x)~.
\labell{cauchyapp}
\eeq
For a proof of this statement see ref.~\cite{symfunc}. This relation is very general and holds for any two abstract
sets of variables $z=\{z_l\}$ and $x=\{x_i\}$. We will only need the case $n=m$ where the representations
$\lambda$ are general frames with $n$ boxes.


\begin{thebibliography}{99}

\bibitem{rolling}
{A.~Sen,
%``Rolling tachyon,''
JHEP {\bf 0204}, 048 (2002)
[arXiv:hep-th/0203211];
%%CITATION = HEP-TH 0203211;%%
%\bibitem{tach1}
%``Tachyon matter,''
JHEP {\bf 0207}, 065 (2002)
[arXiv:hep-th/0203265];
%%CITATION = HEP-TH 0203265;%%
%\bibitem{tach2}
%``Field theory of tachyon matter,''
Mod.\ Phys.\ Lett.\ A {\bf 17}, 1797 (2002)
[arXiv:hep-th/0204143].
%%CITATION = HEP-TH 0204143;%%
}

\bibitem{opensbrane}
{A.~Strominger,
%``Open string creation by S-branes,''
[arXiv:hep-th/0209090].
%%CITATION = HEP-TH 0209090;%%
}

\bibitem{lnt}
{F.~Larsen, A.~Naqvi and S.~Terashima,
%``Rolling tachyons and decaying branes,''
JHEP {\bf 0302}, 039 (2003)
[arXiv:hep-th/0212248].
%%CITATION = HEP-TH 0212248;%%
}

\bibitem{GS}
{M.~Gutperle and A.~Strominger,
%``Timelike boundary Liouville theory,''
[arXiv:hep-th/0301038].
%%CITATION = HEP-TH 0301038;%%
}


\bibitem{sthermo}
{A.~Maloney, A.~Strominger and X.~Yin,
%``S-brane thermodynamics,''
[arXiv:hep-th/0302146];\\
%%CITATION = HEP-TH 0302146;%%
N.~Lambert, H.~Liu and J.~Maldacena,
%``Closed strings from decaying D-branes,''
[arXiv:hep-th/0303139];\\
%%CITATION = HEP-TH 0303139;%%
D.~Gaiotto, N.~Itzhaki and L.~Rastelli,
%``Closed strings as imaginary D-branes,''
[arXiv:hep-th/0304192].
%%CITATION = HEP-TH 0304192;%%
}

%\cite{Sen:2003bc}
\bibitem{moresen}
A.~Sen,
%``Open and closed strings from unstable D-branes,''
[arXiv:hep-th/0207105];
[arXiv:hep-th/0209122];
[arXiv:hep-th/0303057];
[arXiv:hep-th/0305011] .
%%CITATION = HEP-TH 0305011;%%

\bibitem{sbranes}
{M.~Gutperle and A.~Strominger,
%``Spacelike branes,''
JHEP {\bf 0204}, 018 (2002)
[arXiv:hep-th/0202210].
%%CITATION = HEP-TH 0202210;%%
}


\bibitem{moresbrane}
{C.~M.~Chen, D.~V.~Gal'tsov and M.~Gutperle,
%``S-brane solutions in supergravity theories,''
Phys.\ Rev.\ D {\bf 66}, 024043 (2002)
[arXiv:hep-th/0204071];\\
%%CITATION = HEP-TH 0204071;%%
M.~Kruczenski, R.~C.~Myers and A.~W.~Peet,
%``Supergravity S-branes,''
JHEP {\bf 0205}, 039 (2002)
[arXiv:hep-th/0204144];\\
%%CITATION = HEP-TH 0204144;%%
A.~Buchel, P.~Langfelder and J.~Walcher,
%``Does the tachyon matter?,''
Annals Phys.\  {\bf 302}, 78 (2002)
[arXiv:hep-th/0207235]; A.~Buchel and J.~Walcher,
%%CITATION = HEP-TH 0207235;%%
%``The tachyon does matter,''
[arXiv:hep-th/0212150]; [arXiv:hep-th/0305055];\\
%%CITATION = HEP-TH 0212150;%%
F.~Leblond and A.~W.~Peet,
%``SD-brane gravity fields and rolling tachyons,''
[arXiv:hep-th/0303035]; [arXiv:hep-th/0305059];\\
%%CITATION = HEP-TH 0303035;%%
K.~Hashimoto, P.~M.~Ho, S.~Nagaoka and J.~E.~Wang,
%``Time evolution via S-branes,''
[arXiv:hep-th/0303172].
%%CITATION = HEP-TH 0303172;%%
}



\bibitem{minahan}
{J.~A.~Minahan,
%``Rolling the tachyon in super BSFT,''
JHEP {\bf 0207}, 030 (2002)
[arXiv:hep-th/0205098].
%%CITATION = HEP-TH 0205098;%%
}

\bibitem{terashima}
{S.~Sugimoto and S.~Terashima,
%``Tachyon matter in boundary string field theory,''
JHEP {\bf 0207}, 025 (2002)
[arXiv:hep-th/0205085].
%%CITATION = HEP-TH 0205085;%%
}


\bibitem{okuda}
{T.~Okuda and S.~Sugimoto,
%``Coupling of rolling tachyon to closed strings,''
Nucl.\ Phys.\ B {\bf 647}, 101 (2002)
[arXiv:hep-th/0208196].
%%CITATION = HEP-TH 0208196;%%
}

\bibitem{rey1}
{S.~J.~Rey and S.~Sugimoto,
%``Rolling tachyon with electric and magnetic fields: T-duality approach,''
Phys.\ Rev.\ D {\bf 67}, 086008 (2003)
[arXiv:hep-th/0301049];
%%CITATION = HEP-TH 0301049;%%
%``Rolling of modulated tachyon with gauge flux and emergent fundamental  string,''
[arXiv:hep-th/0303133].
%%CITATION = HEP-TH 0303133;%%
}

\bibitem{DaC}
{B.~C.~Da Cunha and E.~J.~Martinec,
%``Closed string tachyon condensation and worldsheet inflation,''
[arXiv:hep-th/0303087];\\
A.~Strominger and T.~Takayanagi,
%``Correlators in timelike bulk Liouville theory,''
[arXiv:hep-th/0303221].
%%CITATION = HEP-TH 0303221;%%
}

\bibitem{Kut}
{D.~Kutasov and V.~Niarchos,
%``Tachyon effective actions in open string theory,''
[arXiv:hep-th/0304045];\\
%%CITATION = HEP-TH 0304045;%%
K.~Okuyama,
%``Wess-Zumino term in tachyon effective action,''
[arXiv:hep-th/0304108].
}

\bibitem{mv}
{J.~McGreevy and H.~Verlinde,
%``Strings from tachyons: The c = 1 matrix reloaded,''
[arXiv:hep-th/0304224].\\
%%CITATION = HEP-TH 0304224;%%
I.~Klebanov, J.~Maldacena and N.~Seiberg. [arXiv:hep-th/0305159].
}


\bibitem{wati1}
{D.~J.~Gross and W.~I.~Taylor,
%``Two-dimensional QCD is a string theory,''
Nucl.\ Phys.\ B {\bf 400}, 181 (1993)
[arXiv:hep-th/9301068].
%%CITATION = HEP-TH 9301068;%%
}

\bibitem{wati2}
{D.~J.~Gross and W.~I.~Taylor,
%``Twists and Wilson loops in the string theory of two-dimensional QCD,''
Nucl.\ Phys.\ B {\bf 403}, 395 (1993)
[arXiv:hep-th/9303046].
%%CITATION = HEP-TH 9303046;%%
}

\bibitem{polchinski}
{J.~Polchinski,
``String Theory, vol. 1'', Cambridge University Press, 2000
}

\bibitem{reptheory}
{C.~W.~ Curtis and I.~Reiner, ``Methods of Representation Theory'',
J. Wiley and Sons, 1981}.

\bibitem{cklm}
{C.~G.~Callan, I.~R.~Klebanov, A.~W.~Ludwig and J.~M.~Maldacena,
%``Exact solution of a boundary conformal field theory,''
Nucl.\ Phys.\ B {\bf 422}, 417 (1994)
[arXiv:hep-th/9402113].
%%CITATION = HEP-TH 9402113;%%
}
\bibitem{RS}
{A.~Recknagel and V.~Schomerus,
%``Boundary deformation theory and moduli spaces of D-branes,''
Nucl.\ Phys.\ B {\bf 545}, 233 (1999)
[arXiv:hep-th/9811237].
%%CITATION = HEP-TH 9811237;%%
}

\bibitem{fzz}
{V.~Fateev, A.~B.~Zamolodchikov and A.~B.~Zamolodchikov,
%``Boundary Liouville field theory. I: Boundary state and boundary  two-point function,''
[arXiv:hep-th/0001012].
%%CITATION = HEP-TH 0001012;%%
}

\bibitem{schurweyl}
{W.~J.~Holman and L.~C.~Biedenharn, ``The Representations and Tensor Operators
of the Unitary Groups,'' in ``Group Theory and its Applications'' ed. E.~M.~Loebl,
Academic Press, 1971}.

\bibitem{weyl}
{H.~Weyl, ``Group Theory and Quantum Mechanics'', Dover Press.
}

\bibitem{symfunc}
{I.~G.~Macdonald, ``Symmetric Functions and Hall Polynomials, 2nd Edition'', 
Oxford University Press, 1995.\\
L.~Manivel, ``Symmetric Functions, Schubert Polynomials and Degeneracy Loci'',
American Mathematical Society Monogrpahs, 2001.
}

\bibitem{maple}
{S. Veigneau, ``ACE, an Algebraic Combinatorics Environment for the computer 
algebra system MAPLE'',  http://phalanstere.univ-mlv.fr/~ace/
}






\end{thebibliography}
\end{document}